\documentclass[%
aps, 
pre,
amsmath,amssymb,
reprint,%
]{revtex4-1}
\usepackage{graphicx}
\usepackage{dcolumn}
\usepackage{bm}

\usepackage{amsmath}	
\begin{document}

\newcommand{\diff}[2]{\frac{d#1}{d#2}}
\newcommand{\pdiff}[2]{\frac{\partial #1}{\partial #2}}
\newcommand{\fdiff}[2]{\frac{\delta #1}{\delta #2}}
\newcommand{\new}{\nonumber\\}
\newcommand{\bx}{\bm{x}}
\newcommand{\by}{\bm{y}}
\newcommand{\bu}{\bm{u}}
\newcommand{\hr}{\hat{r}}
\newcommand{\hx}{\hat{x}}
\newcommand{\hf}{\hat{f}}
\newcommand{\hpi}{\hat{\pi}}
\newcommand{\bX}{\bm{X}}
\newcommand{\tq}{\tilde{q}}
\newcommand{\tD}{\tilde{D}}
\newcommand{\de}{\mathrm{d}}
\newcommand{\tk}{\tilde{k}}
\newcommand{\tmu}{\tilde{\mu}}
\newcommand{\txi}{\tilde{\xi}}
\newcommand{\teta}{\tilde{\eta}}
\newcommand{\abs}[1]{\left|#1\right|}
\newcommand{\ave}[1]{\left\langle #1\right\rangle}
\newcommand{\M}{\mathcal{M}}
\newcommand{\Y}{\mathcal{Y}}
\newcommand{\T}{\mathcal{T}}
\newcommand{\A}{\mathcal{A}}
\newcommand{\tC}{\tilde{C}}
\newcommand{\Z}{\mathcal{Z}}
\newcommand{\G}{\mathcal{G}}
\newcommand{\PP}{\mathcal{P}}
\newcommand{\he}{\mathcal{H}}
\newcommand{\im}{{\rm Im}}
\newcommand{\erf}{{\rm erf}}
\newcommand{\var}[1]{{\rm Var}\left[#1\right]}
\newcommand{\pr}{{\rm PR}}
\newcommand{\tm}{\tilde{\mu}}

\preprint{AIP/123-QED} \title{Interaction-free ergodicity-breaking
driven by temporally hyperuniform noise}


\author{Harukuni Ikeda}
 \email{harukuni.ikeda@yukawa.kyoto-u.ac.jp}
\affiliation{Yukawa Institute for Theoretical Physics, Kyoto University,
Kyoto 606-8502, Japan}

\date{\today}

\begin{abstract}
We show that norm-conserving spin models driven by temporally
hyperuniform noise exhibit a sharp ergodicity-breaking transition in the
absence of interactions. In the nonergodic phase, the dynamics freeze
into configurations determined by the initial condition. Our analysis
demonstrates that such interaction-free ergodicity breaking arises
generically whenever a global constraint is imposed and the driving
noise is class-I hyperuniform, the strongest form in
Torquato's classification. The transition can also be interpreted as a
condensation of fluctuations into the zero-frequency mode, reminiscent
of Bose--Einstein condensation in an ideal gas.
\end{abstract}

\maketitle

\section{Introduction}

The concept of hyperuniformity was introduced by Torquato and
Stillinger, and characterizes the anomalous suppression of
long-wavelength fluctuations in many-body systems
\cite{torquato2003,torquato2018hyperuniform}. Hyperuniformity is defined
by the power-law scaling of the structure factor, $S(k)\sim |k|^\alpha$
with $\alpha>0$ for small wave vectors $k$. This property has been
reported in a wide range of systems, including
crystals~\cite{kim2018effect},
quasicrystals~\cite{oguz2017,koga2024hyperuniform}, amorphous
solids~\cite{donev2005,hopkins2012,ikeda2015thermal,hexner2018,hexner2019,
bolton2024ideal,wang2025hyperuniform}, and various nonequilibrium
settings~\cite{hexner2017noise,lei2019nonequilibrium,lei2019hydrodynamics,kuroda2023microscopic,leonardo2023,ikeda2023cor,ikeda2023harmonic,maire2024enhancing,kuroda2024long,ikeda2024continuous}. The
influence of hyperuniformity on structural and dynamical properties has
been an active topic of
investigation~\cite{sire1993ising,schwartz1993,chandran2017,sakai2022quantum,luck1993critical,chandran2017,hyperuniformrandom2019,
hexner2017noise,leonardo2023,ikeda2023cor,ikeda2023harmonic,maire2024enhancing}.

A natural generalization of this idea is temporal hyperuniformity. A
process is called temporally hyperuniform if its spectral density
behaves as $\tilde{D}(\omega)\sim |\omega|^\alpha$ at small frequencies
$\omega$ \cite{torquato2018hyperuniform}. Such processes arise in
diverse contexts, including fractional Brownian motion with the Hurst
exponent $H<1/2$ \cite{mandelbrot1968fractional}, stochastic resetting
\cite{evans2011,evans2020}, high-pass filtered noise
\cite{guz1998,bao1999,guz2001}, and avalanche statistics in sandpile
models \cite{GarciaMillan2018}, see also Appendix~\ref{193904_26Jul25}
for a tentative list of analytically tractable examples. Torquato
classified hyperuniform processes into three classes: class I
($\alpha>1$), where the variance converges to a finite value; class II
($\alpha=1$), where it grows logarithmically; and class III
($0<\alpha<1$), where it grows subdiffusively.

Previous studies have explored the impact of class-III temporal
hyperuniform noise on phase transitions
\cite{ikeda2023cor,ikeda2023harmonic}. In particular, anticorrelations
in the noise were shown to reduce the lower critical dimension, enabling
continuous symmetry breaking even in one and two dimensions, which is
forbidden in equilibrium by the Hohenberg–Mermin–Wagner theorem
\cite{hohenberg1967,mermin1966}. By contrast, the effects of class I and
II hyperuniform noise on many-body systems remain largely unexplored.

In this work, we address this gap by considering a minimal toy model: a
noninteracting system subject only to a global norm-conservation
constraint. Such constrained systems have been widely studied in
statistical mechanics, quantum mechanics, neural networks, and so
on~\cite{sakurai2020modern,berlin1952spherical,stanley1968,gardner1988space,crisanti1992sphericalp}. In
the absence of interactions, these models are not expected to undergo
phase transitions. Remarkably, we show that this expectation breaks down
when class-I temporally hyperuniform noise drives the system. Even
without interactions, the system undergoes an ergodicity-breaking
transition. The transition can be interpreted as a condensation of
fluctuations into the zero-frequency mode, analogous to Bose–Einstein
condensation (BEC). Unlike conventional BEC, where the ground state is
energetically favored, all configurations in our model are degenerate,
and the condensed state is selected dynamically rather than
thermodynamically. We further generalize our analysis to conserved $L_p$
norms as well as models with soft constraints, and demonstrate that
ergodicity breaking generically emerges under class-I hyperuniform
driving.

The remainder of this paper is organized as follows. In
Sec.~\ref{184817_6Mar25}, we introduce the
model. Sec.~\ref{174550_29Jul25} presents numerical evidence for
ergodicity breaking, while Sec.~\ref{175223_29Jul25} provides analytical
results from dynamical mean-field theory. In Sec.~\ref{175652_29Jul25},
we analyze the dependence on the noise
spectrum. Sec.~\ref{112809_27Jul25} compares the ergodicity breaking
with Bose–Einstein condensation in the ideal Bose gas and the
ferromagnetic transition in mean-field spin models. In
Sec.~\ref{183622_29Jul25}, we generalize the analysis to models with
conserved $L_p$ norms for arbitrary $p$ as well as models with soft
constraints. Finally, Sec.~\ref{164632_2Sep25} concludes the paper.

\section{Model}
\label{184817_6Mar25}

We consider a system of $N$ continuous spin variables
$\{\sigma_1(t),\cdots, \sigma_N(t)\}$
with the fixed norm:
\begin{align}
\sum_{i=1}^N \sigma_i(t)^2 =N.\label{162436_18Jul25}
\end{align}
Norm conservation arises in diverse fields, including quantum mechanics,
statistical mechanics, machine learning, and so
on~\cite{sakurai2020modern,berlin1952spherical,stanley1968,kirkpatrick1987,crisanti1992sphericalp,castellani2005spin,castellani2005spin,cavagna2009supercooled,ikeda2023cor,ikeda2024continuous,gardner1988space,gardner1988optimal,franz2017universality}. In
statistical mechanics, the continuous spin model with the
constraint~(\ref{162436_18Jul25}) is known as the spherical model that
has been extensively studied as a solvable model for various phase
transitions, including the ferromagnetic
transition~\cite{berlin1952spherical,castellani2005spin}, glass
transition~\cite{kirkpatrick1987p,castellani2005spin}, and jamming
transition~\cite{franz2016,franz2017universality}. In equilibrium, phase
transitions in the spherical model are driven by spin–spin
interactions. In contrast, here we report a novel type of
out-of-equilibrium phase transition that occurs in the absence of
interactions, induced solely by the strong anticorrelation of noise.  The
dynamics of each spin are governed by the following equation of motion
(EOM):
\begin{align}
\dot{\sigma}_i(t) 
 = -\mu(t)\sigma_i(t) + \xi_i(t),\label{115517_4Mar25}
\end{align}
where $\mu(t)$ denotes the Lagrange multiplier to impose the
constraint~(\ref{162436_18Jul25}).  From the condition
$\frac{d}{dt}\sum_{i=1}^N \sigma_i^2=0$, 
$\mu(t)$ is determined as 
\begin{align}
\mu(t) = \frac{1}{N}\sum_{i=1}^N \sigma_i(t) \xi_i(t).
\end{align}
The noise $\xi_i(t)$ is Gaussian 
with zero mean and variance: 
\begin{align}
\ave{\xi_i(t)\xi_j(t')}  = 2\delta_{ij}TD(t),
\end{align}
where $T$ denotes the noise amplitude. For thermal white noise, $D(t)$
reduces to a delta function, yielding a flat Fourier spectrum.  In
contrast, this work focuses on temporally hyperuniform noise,
characterized by a Fourier spectrum
\begin{align}
\tD(\omega) \propto \abs{\omega}^\alpha,\quad \alpha>0
\end{align}
for small $\omega$. This scaling implies that long-time fluctuations are
highly suppressed, {\it i.e.}, the noise exhibits temporal
hyperuniformity.

\section{Numerical simulation}
\label{174550_29Jul25} We first present a numerical demonstration of the
interaction-free ergodicity breaking of the spherical model driven by
noise generated by a high-pass filter, which provides one of the simplest
realizations of temporally hyperuniform noise.
\subsection{Protocol to generate temporally hyperuniform noise}
A simple and efficient way to generate temporally hyperuniform noise is
to apply a high-pass filter to Gaussian white noise.  Let $\eta_i(t)$ 
denote Gaussian white noise with zero mean and variance
\begin{align}
\ave{\eta_i(t)\eta_j(t')} = 2T\delta_{ij}\delta(t-t').
\end{align}
The high-pass filtered noise $\xi_i(t)$ is obtained by solving the
following equation of motion~\cite{guz1998,guz2001}:
\begin{align} 
&\xi_i(t) =  \dot{x}_i(t),
&\dot{x}_i(t) = -kx_i(t)+\eta_i(t).\label{195400_20Jul25}
\end{align}
The noise can also be identified as velocity fluctuations of a Brownian
particle in a harmonic trap~\cite{bao2005,hu2017,chen2018,li2021}, see
Appendix~\ref{highpass} and \ref{harmonic}. In the steady state, the
noise correlation is
\begin{align}
\ave{\xi_i(t)\xi_j(t')} = 2T\delta_{ij}D(t-t'),
\end{align}
with the Fourier transform  
\begin{align}
\tD(\omega) \equiv 
\int_{-\infty}^{\infty} dt e^{i\omega t}D(t)
 = \frac{\omega^2}{\omega^2+k^2}.\label{124131_21Jul25}
\end{align}
For $\omega \ll 1$, one finds $\tD(\omega)\propto \omega^2$, which
vanishes in the limit $\omega \to 0$. This implies that fluctuations of
$\xi_i(t)$ are strongly suppressed at long time scales.  Indeed, the
mean-squared displacement (MSD) of the integrated noise converges to a
finite value for large $t$:
\begin{align}
\ave{\left(\int_0^t \xi_i(t)\right)^2}
= \frac{2T(1-e^{-kt})}{k}
\xrightarrow{t\to\infty} \frac{2T}{k},
\end{align}
meaning that the noise is class-I
hyperuniform~\cite{torquato2018hyperuniform}. 

\subsection{Discretization}
For the numerical integration, we discretize the
EOMs~(\ref{115517_4Mar25}) and (\ref{195400_20Jul25}) while preserving
the spherical constraint~(\ref{162436_18Jul25}).  
Following Ref.~\cite{hwang2020}, we update the spins as 
\begin{align}
\sigma_i(t+\Delta t) =
\frac{\sigma_i'(t+\Delta t)}
 {\sqrt{N^{-1}\sum_{i=1}^N \sigma_i'(t+\Delta t)^2}},\label{200314_20Jul25}
\end{align}
where 
\begin{align}
\sigma_i'(t+\Delta t) =  \sigma_i(t) +  x_i(t+\Delta t)-x_i(t),
\end{align}
and 
\begin{align}
x_i(t+\Delta t) = x_i(t) -k x_i(t)\Delta t + \sqrt{2T\Delta t}W_t.
\end{align}
Here, $W_t$ is an independent and identically distributed random
variable of zero mean and unit variance:
 \begin{align}
& \ave{W_t}=0,
&\ave{W_tW_{t'}}=\delta_{t,{t'}}.
 \end{align}
Eq.~(\ref{200314_20Jul25}) explicitly preserves the spherical constraint
$\sum_{i=1}^N \sigma_i^2=N$, and the original EOMs are recovered in the
continuum limit $\Delta t\to 0$. Note that these equations do not
involve any explicit interactions. Nevertheless, the model exhibits a
sharp ergodicity-breaking transition, as we will see below.

To characterize the ergodicity, 
we measure the steady-state time correlation function:
\begin{align}
C(t) = \lim_{t_w \to\infty }C(t+t_w,t_w), 
\end{align}
where
\begin{align}
C(t,t') = \frac{1}{N}\sum_{i=1}^N \ave{\sigma_i(t)\sigma_i(t')}.
\end{align}
Numerical simulations were performed for ${t_w=10^2}$, ${N=10^5}$, and
${\Delta t=10^{-2}}$, for which convergence was carefully verified.

\subsection{Results}

\begin{figure}[t]
\begin{center}
\includegraphics[width=8cm]{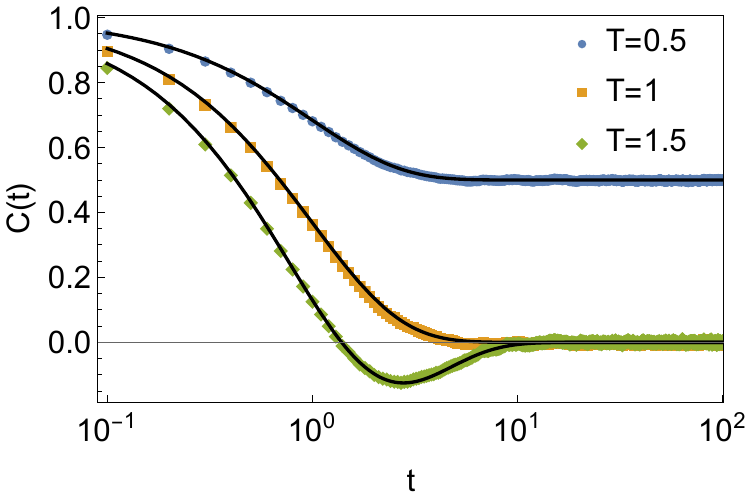} \caption{ Correlation function
$C(t)$ of the spherical model driven by high-pass filtered noise.
Markers denote numerical results, while solid lines show the theoretical
predictions.  For $T=1.5$ and $1.0$, $C(t)$ decays to zero at long
times, indicating ergodicity. For $T=0.5$, $C(t)$ converges to a finite
value, indicating nonergodicity.}  \label{213759_20Jul25}
\end{center}
\end{figure}
Fig.~\ref{213759_20Jul25} shows $C(t)$ for several $T$ with
$k=1$. For large $T$ ($T=1.5, 1.0$), $C(t)$ decays
to zero, whereas for small $T$ ($T=0.5$), it converges to a finite value.
These results suggest the existence of a nonergodic transition 
at a critical value of $T$.
\begin{figure}[t]
\begin{center}
\includegraphics[width=8cm]{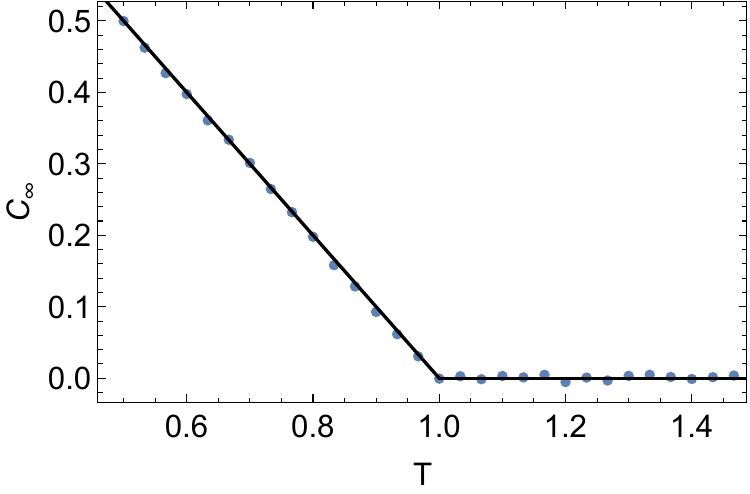} \caption{ $T$ dependence of
nonergodicity parameter $C_\infty=\lim_{t\to\infty}C(t)$ of the
spherical model driven by high-pass filtered noise for $k=1$. Markers
denote numerical results, while solid line shows theoretical
prediction. } \label{112410_21Jul25}
\end{center}
\end{figure}
Fig.~\ref{112410_21Jul25} shows the $T$ dependence of the
nonergodicity parameter:
\begin{align}
C_\infty\equiv \lim_{t\to\infty}C(t).\label{123413_28Jul25}
\end{align}
For $T>1$, $C_\infty\approx 0$, while for $T<1$, $C_\infty$ remains
finite, indicating ergodicity breaking at $T=1$. This is a rather
surprising result given that the model contains no explicit
interactions. Understanding the mechanism behind this phenomenon is the
central aim of this work.

\section{Dynamical mean-field theory}
\label{175223_29Jul25} We now develop a dynamical mean-field theory to
calculate the steady-state correlation and response functions of the model.

\subsection{Correlation and response}

The correlation and response functions are defined
as~\cite{castellani2005spin}
\begin{align}
C(t,t') &= \frac{1}{N}\sum_{i=1}^N \ave{\sigma_i(t)\sigma_i(t')},\\ 
R(t,t') &= \frac{1}{N}\sum_{i=1}^N \ave{\fdiff{\sigma_i(t)}{\xi_i(t')}}.
\end{align}
Multiplying both sides of Eq.~(\ref{115517_4Mar25})
by $\sigma_i(t')$ 
and summing over $i$, 
we obtain~\cite{castellani2005spin}
\begin{align}
\pdiff{C(t,t')}{t}
= -\mu(t) C(t,t')
 +2T\int_{-\infty}^{\infty} ds D(t-s)R(t',s),\label{175823_14Jun25}
\end{align}
where we used the identity obtained from the generating functional
formalism~\cite{msr1973}:
\begin{align}
\frac{1}{N}\sum_{i=1}^N \ave{\sigma_i(t)\xi_i(t')}  &=
\frac{1}{N}\sum_{i=1}^N \int_{-\infty}^{\infty}ds
 \ave{\fdiff{\sigma_i(t)}{\xi_i(s)}}\ave{\xi_i(s)\xi_i(t')}\new 
&=2T\int_{-\infty}^{\infty} ds D(t-s)R(t',s).
\end{align}
Taking the functional derivative of Eq.~(\ref{115517_4Mar25}) with
respect to $\xi_i(t')$ and summing over $i$, we obtain
\begin{align}
\pdiff{R(t,t')}{t} = -\mu(t)R(t,t') +\delta(t-t').\label{173510_14Jun25}
\end{align}
The Lagrange multiplier $\mu(t)$ is determined by the condition of the
norm-conservation $C(t,t)=1$, which leads to
\begin{align}
\left[\pdiff{C(t,t')}{t} + \pdiff{C(t,t')}{t'} \right]_{t=t'}
= 0.
\end{align}
From the above equation, after some manipulations, 
we get~\cite{castellani2005spin}
\begin{align}
\mu(t) = 2T\int dt' D(t-t')R(t,t').
\end{align}

\subsection{Steady-state solution}

In the steady state, the system exhibits 
time-translation invariance, 
so that 
\begin{align}
C(t,t') = C(t-t'),\qquad R(t,t') = R(t-t').
\end{align}
Eq.~(\ref{175823_14Jun25})
then reduces to 
\begin{align}
\pdiff{C(t)}{t} = -\mu C(t) + 2T\int_{-\infty}^{\infty}dt'D(t+t')R(t').\label{190835_26Jun25}
\end{align}
The equation for the response function~(\ref{173510_14Jun25})
reduces to 
\begin{align}
\pdiff{R(t)}{t} = -\mu R(t) + \delta(t),
\end{align}
which has the solution
\begin{align}
R(t) = \theta(t)e^{-\mu t}.\label{190840_26Jun25} 
\end{align}
Stability requires $\mu\geq 0$,
since otherwise $R(t)$ would diverge at long times.
The Lagrange
multiplier $\mu$ is determined by the self-consistent equation
\begin{align}
\mu &= 2T\int_{-\infty}^{\infty}dt D(t)R(t)\new
 &= \frac{2T\mu}{\pi}\int_0^\infty d\omega
 \frac{\tD(\omega)}{\mu^2+\omega^2}.\label{180124_14Jun25}
\end{align}
Eq.~(\ref{180124_14Jun25}) admits the trivial solution $\mu=0$.
Assuming $\mu>0$, a non-trivial solution is obtained from the
self-consistent equation
\begin{align}
 1 = \frac{2T}{\pi}\int_0^\infty d\omega
 \frac{\tD(\omega)}{\mu^2+\omega^2}.\label{002306_7Jul25}
\end{align}
Using Eqs.~(\ref{190835_26Jun25}) and (\ref{190840_26Jun25}), the
correlation function can be written as
\begin{align}
C(t) &= C(0)e^{-\mu t} + 2T\int_0^t dt_1 e^{-\mu(t-t_1)}
 \int_0^{\infty} dt_2 D(t_1+t_2)e^{-\mu t_2}\new
&= e^{-\mu t}
\left(
1 -\frac{2T}{\pi}\int_0^\infty d\omega
 \frac{\tD(\omega)}{\omega^2+\mu^2} \right)\new 
&+ \frac{2T}{\pi}\int_0^\infty d\omega
 \frac{\tD(\omega)}{\omega^2+\mu^2}\cos(\omega t).
\end{align}
For $\mu>0$, 
substituting Eq.~(\ref{002306_7Jul25}) into the above equation, 
we obtain 
\begin{align}
C(t) =  \frac{2T}{\pi}\int_0^\infty d\omega
 \frac{\tD(\omega)}{\omega^2+\mu^2}\cos(\omega t),\label{132018_21Jul25}
\end{align}
which decays to zero as $t\to\infty$ due to the oscillatory cosine
factor.  In contrast, for $\mu=0$, we obtain 
\begin{align}
 C(t) = 1 -\frac{2T}{\pi}\int_0^\infty d\omega
 \frac{\tD(\omega)}{\omega^2}
 + \frac{2T}{\pi}\int_0^\infty d\omega
 \frac{\tD(\omega)}{\omega^2}\cos(\omega t).\label{122827_21Jul25}
\end{align}
The last term vanishes in the long-time limit, 
yielding the nonergodicity parameter
\begin{align}
C_{\infty}  = 1 -\frac{2T}{\pi}\int_0^\infty d\omega
\frac{\tD(\omega)}{\omega^2}.\label{130200_21Jul25}
\end{align}
At the transition point $T=T_c$, $C_{\infty}=0$, 
leading to 
\begin{align}
 T_c = \frac{\pi}{2}
\left[ \int_0^\infty d\omega
\frac{\tD(\omega)}{\omega^2}\right]^{-1}.\label{124448_21Jul25}
\end{align}
In summary, for $T>T_c$, the system is ergodic, where
the correlation function is given by
Eq.~(\ref{132018_21Jul25}), whereas for $T<T_c$, the system is nonergodic,
where the correlation function is given by Eq.~(\ref{122827_21Jul25}).

\subsection{Comparison with numerical simulation}
\label{highpass} We now apply the above framework to the case of
high-pass filtered white noise. Substituting Eq.~(\ref{124131_21Jul25})
into Eq.~(\ref{124448_21Jul25}), we obtain the transition point
\begin{align}
T_c = k.
\end{align}
For $T<T_c$, the nonergodicity parameter~(\ref{130200_21Jul25}) is
written as
\begin{align}
C_{\infty} = 1-\frac{T}{T_c}.\label{125135_23Jul25}
\end{align}
In Fig.~\ref{112410_21Jul25}, we show the above theoretical prediction
with the black solid line, which agrees well with the numerical results.

For $T>T_c$, the steady-state correlation
function is obtained from Eq.~(\ref{132018_21Jul25}) as 
\begin{align}
C(t)= \frac{T(ke^{-kt}-\mu e^{-\mu t})}{k^2-\mu^2}.\label{132727_21Jul25}
\end{align}
Here, the Lagrange multiplier $\mu$ is determined by the self-consistent
equation~(\ref{002306_7Jul25}), which leads to
\begin{align}
 1 = \frac{T}{k+\mu} \to \mu = T-T_c.\label{121404_22Jul25}
\end{align}
For $T<T_c$, Eq.~(\ref{122827_21Jul25}) gives 
\begin{align}
 C(t) = C_\infty+ \frac{T}{k}e^{-kt}.\label{132722_21Jul25}
\end{align}
In Fig.~\ref{213759_20Jul25}, we show the theoretical predictions for
$C(t)$, Eqs.~(\ref{132727_21Jul25}) and (\ref{132722_21Jul25}), with the
black solid lines, which agree well with the numerical results,
validating our theoretical calculations.

\section{Power spectrum dependence}
\label{175652_29Jul25}

Here, we investigate how the strength of hyperuniformity affects
ergodicity breaking. In particular, we show that the transition occurs
only when the driving noise is class-I hyperuniform.

\subsection{Phase diagram}
\begin{figure}[t]
\begin{center}
\includegraphics[width=8cm]{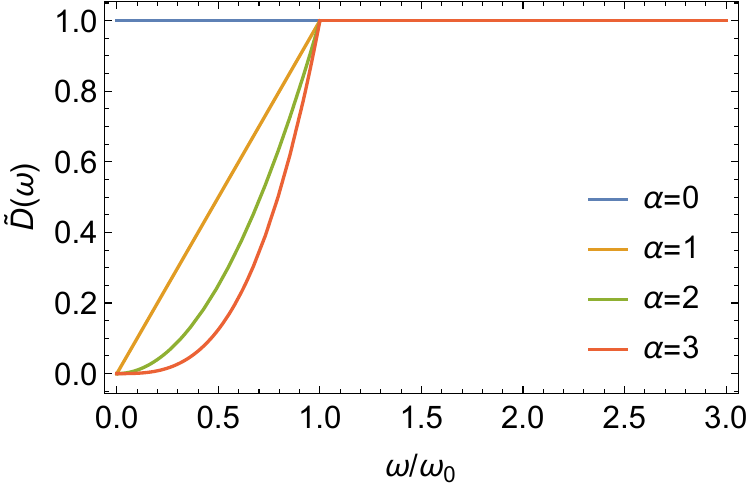} \caption{ $\tD(\omega)$ for several
values of $\alpha$.  The case $\alpha=0$ corresponds to thermal white
noise.  For $\alpha>0$, $\tD(\omega)$ vanishes in the limit $\omega\to
0$.}  \label{163621_6Mar25}
\end{center}
\end{figure}

To systematically investigate the
nonergodic transition induced by temporally hyperuniform noise, we
consider the following spectrum:
\begin{align}
\tD(\omega) 
=
 \begin{cases}
 \abs{\omega/\omega_0}^\alpha & \abs{\omega} \leq \omega_0\\
 1 & \abs{\omega} > \omega_0
 \end{cases},\label{190802_18Jul25}
\end{align}
where $\alpha$ and $\omega_0$ are positive constants. As shown in
Fig.~\ref{163621_6Mar25}, $\tD(\omega)$ vanishes at small $\omega$,
indicating that the noise is temporally hyperuniform. The white noise is
recovered in the limit $\omega_0\to 0$ or $\alpha\to 0$. The noise can
be classified into three classes by examining the variance of the
integrated noise~\cite{torquato2018hyperuniform}, which corresponds to
the MSD of a free particle driven by the noise. For $\alpha > 1$, the
MSD saturates at long times, indicating class-I hyperuniformity; for
$\alpha = 1$, it grows logarithmically, indicating class-II
hyperuniformity; and for $0 < \alpha < 1$, it grows sublinearly,
indicating class-III hyperuniformity (see Appendix~\ref{180030_11Sep25}
for detailed calculations).

\begin{figure}[t]
\begin{center}
\includegraphics[width=8cm]{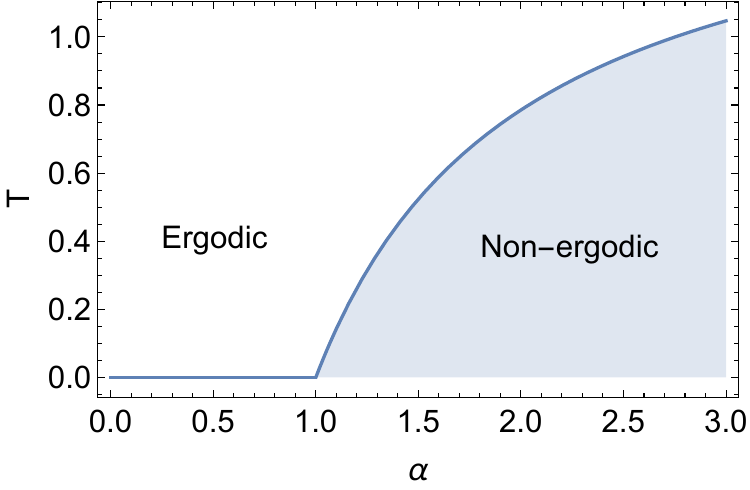} \caption{ Phase diagram for
$\omega_0=1$. Solid line denotes the transition line $T_c$, and shaded
region indicates the nonergodic phase. } \label{154045_21Jul25}
\end{center}
\end{figure}
The transition point $T_c$ can be obtained by substituting
Eq.~(\ref{190802_18Jul25}) into Eq.~(\ref{124448_21Jul25}), 
leading to 
\begin{align}
T_c =
\begin{cases}
 \frac{\pi\omega_0}{2}\frac{\alpha-1}{\alpha} & \alpha > 1 \\ 
0 & \alpha \leq 1
\end{cases}.
\end{align}
The resulting phase
diagram is shown in Fig.~\ref{154045_21Jul25}. The model undergoes a
nonergodic transition at finite $T_c$ for $\alpha>1$.  In other words,
when the noise is class-I temporal hyperuniform, strong noise
anticorrelation alone can induce the nonergodic transition in the
absence of interactions.

\subsection{Scaling behavior}

Eq.~(\ref{190840_26Jun25}) implies that the response function
decays as $R(t)=\theta(t)e^{-t/\tau}$ with the relaxation time
\begin{align}
\tau=1/\mu.
\end{align}
The Lagrange multiplier $\mu$ is determined by the self-consistent
equation~(\ref{002306_7Jul25}).  The asymptotic analysis near $T_c$
yields~\footnote{For the details of calculations, see, for instance,
Appendix~A in Ref.~\cite{ikeda2023cor}.}
\begin{align}
 \mu \propto
 \begin{cases}
  (T-T_c)^{\frac{1}{\alpha-1}} & 1<\alpha < 3 \\ 
  (T-T_c)^{\frac{1}{2}} & \alpha > 3,
 \end{cases}\label{135152_23Jul25}
\end{align}
and therefore 
\begin{align}
\tau \propto   \begin{cases}
  (T-T_c)^{-\frac{1}{\alpha-1}} & 1<\alpha < 3 \\ 
  (T-T_c)^{-\frac{1}{2}} & \alpha > 3.
 \end{cases}\label{182511_12Aug25}
\end{align}
The critical exponent of the relaxation time depends continuously on
$\alpha$ for $1<\alpha<3$, and saturates to a constant value for
$\alpha>3$. 

The above behaviors are qualitatively analogous to
the dimensional dependence
in conventional critical
phenomena~\cite{nishimori2010elements}: the transition does not occur
below the lower critical dimension $d_{\rm low}$, and the critical
exponents do not depend on the spatial dimensions above the upper
critical dimension $d_{\rm up}$. In the present model, $\alpha_{\rm
low}=1$ plays the role of the lower critical dimension, whereas
$\alpha_{\rm up}=3$ corresponds to the upper critical dimension.

\section{Comparison with other transitions}
\label{112809_27Jul25}

We now compare the ergodicity breaking in our model with other types of
phase transitions, namely, Bose-Einstein condensation and
ferromagnetic transition.

\subsection{Comparison with Bose-Einstein condensation}
The ergodicity breaking in our model arises from a mechanism closely
analogous to Bose–Einstein condensation (BEC), but occurring in the
frequency domain. To see this, it is convenient to solve
Eq.~(\ref{115517_4Mar25}) by Fourier transformation:
\begin{align}
\tilde{\sigma}_i(\omega) &= \frac{\tilde{\xi}_i(\omega)}{i\omega + \mu},
\end{align}
where we assumed that the system is in the steady state, so that $\mu$ is
time independent. The two-point correlation function is then
\begin{align}
\ave{\tilde{\sigma}_i(\omega)\tilde{\sigma}_i(\omega')}  = 2\pi\delta(\omega+\omega')
\tilde{C}_i(\omega),
\end{align}
where
\begin{align}
\tilde{C}_i(\omega) = \int dt e^{i\omega t}\ave{\sigma_i(t)\sigma_i(0)}
 = \frac{2T\tD(\omega)}{\omega^2 + \mu^2}.
\end{align}
The equal-time correlation is given by 
\begin{align}
\ave{\sigma_i(t)^2} &= \frac{1}{2\pi}
 \int_{-\infty}^{\infty} d\omega \tilde{C}_i(\omega)\new
&= \frac{2T}{\pi} \int_0^\infty d\omega \frac{\tD(\omega)}{\omega^2+\mu^2}.
\end{align}
The spherical constraint $N^{-1}\sum_{i=1}^N\ave{\sigma_i(t)^2}=1$ 
thus reduces to 
\begin{align}
1 = TF(\mu),\label{182519_22Jul25}
\end{align}
with 
\begin{align}
F(\mu) = \frac{2}{\pi}\int_0^\infty d\omega 
\frac{\tD(\omega)}{\omega^2+\mu^2}.\label{170911_27Jul25}
\end{align}
This is identical to Eq.~(\ref{002306_7Jul25}), 
and determines $\mu$. Since $F(\mu)$ is monotonically decreasing
and takes its maximum at $\mu=0$,
Eq.~(\ref{182519_22Jul25}) has no solution for $T<T_c=
F(0)^{-1}$, which is the signature of a condensation transition~\cite{greiner2012thermodynamics}.
In this case, $\sigma_i(t)$ must be separated as 
\begin{align}
\sigma_i(t) = \sigma_i^0 + \delta \sigma_i(t),
\end{align}
where $\delta \sigma_i(t)$ denotes the time-dependent component 
that relaxes to zero at long times, 
and $\sigma_i^0$ denotes the static component. 
The nonergodicity parameter is then 
\begin{align}
C_{\infty} = \frac{1}{N}\sum_{i=1}^N \ave{\left(\sigma_i^0\right)^2}.
\end{align}
The spherical constraint becomes
\begin{align}
 1 &= \frac{1}{N}\sum_{i=1}^N \ave{\left(\sigma_i^0\right)^2}
 + \frac{1}{N}\sum_{i=1}^N \ave{\delta\sigma_i^2}\new
&\to C_{\infty} = 1- TF(0) = 1-\frac{T}{T_c},
\end{align}
in agreement with the previous result Eq.~(\ref{125135_23Jul25}). As
discussed in the previous section, for a power-law spectrum
$\tD(\omega)\sim \abs{\omega}^\alpha$ for $\omega\ll 1$, $T_c$ is finite
only for $\alpha>1$, and the scaling of $\mu$ is given by
Eqs.~(\ref{135152_23Jul25}).

The condensation in the frequency space has the same mathematical
structure as BEC of ideal Bose
gas~\cite{greiner2012thermodynamics,crisanti2019}. For BEC, the
total particle number is
\begin{align}
N = G(\mu_{\rm B}) = \int dq g(q)n(q,\mu_{\rm B}),\label{153042_24Jul25}
\end{align}
where $\mu_{\rm B}$ denotes the chemical potential, 
$g(q)$ denotes the density of states, 
and $n(q,\mu_{\rm B})$ denotes the Bose distribution:
\begin{align}
n(q,\mu_{\rm B}) = \frac{1}{e^{-\beta\mu_{\rm B}+\frac{\beta\hbar^2q^2}{2m}}-1}.
\end{align}
From Debye theory, $g(q)\propto \abs{q}^{d-1}$
at small $q$, where $d$ denotes the spatial
dimension~\cite{greiner2012thermodynamics}. Since the small-$q$
regime dominates the scaling, 
we split Eq.~(\ref{153042_24Jul25}) as 
\begin{align}
&N = G_1(\mu_{\rm B}) + G_2(\mu_{\rm B}),\\
&G_1(\mu_{\rm B}) \equiv \int_0^{\Delta q} dqg(q)n(q)
 \approx  CT\int_0^{\Delta q}\frac{q^{d-1}dq}{q^2-\frac{2m\mu_{\rm B}}{\hbar^2}},\\
&G_2(\mu_{\rm B}) \equiv
\int_{\Delta q}^{\infty}dq
 g(q)n(q),\label{162010_24Jul25}
\end{align}
where $\Delta q$ denotes a small cutoff, and $C$ denotes a constant. The
scaling of $\mu_{\rm B}$ is governed by $G_1(\mu_{\rm B})$, which is
mathematically equivalent to $F(\mu)$ in Eq.~(\ref{170911_27Jul25}),
under the correspondence
\begin{align}
&\mu_{\rm B} \leftrightarrow -\frac{\hbar^2\mu^2}{2m},
&d \leftrightarrow \alpha+1.
\end{align}
Thus, $\alpha_{\rm low}=1$ and $\alpha_{\rm up}=3$ correspond to the
lower and upper critical dimensions $d_{\rm low}=2$ and $d_{\rm up}=4$,
respectively, consistent with the known results of BEC of the ideal Bose
gas~\cite{gunton1968condensation,crisanti2019}.

In BEC, the ground state is occupied by a macroscopic number of
particles in momentum space. In contrast, in our model, condensation
occurs in frequency space. Moreover, the energy landscape of the model
is completely flat, and thus the ground state is completely degenerate.
Consequently, the frozen-spin configuration in the nonergodic phase is
determined by the initial condition rather than by energetic selection.

\subsection{Comparison with ferromagnetic transition}
\label{181044_29Jul25} 
\begin{figure}[t]
\begin{center}
\includegraphics[width=8cm]{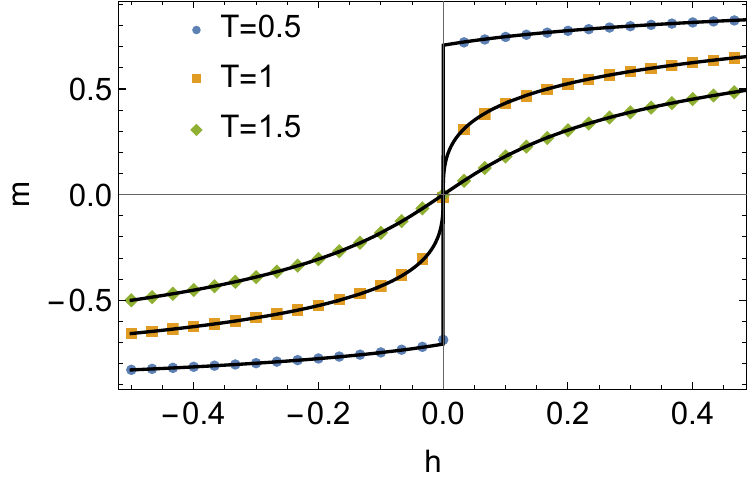} \caption{External field $h$
dependence of the order parameter $m$ for the spherical model driven by
high-pass filtered white noise with $k=1$. Markers denote numerical
results, while solid lines denote theoretical predictions. For $T\geq
T_c=1$, $m$ increases continuously with $h$, whereas for $T<T_c$,
$m$ changes discontinuously at $h=0$. } \label{173331_28Jul25}
\end{center}
\end{figure}

So far, ergodicity breaking has been
characterized only by the dynamical order parameter defined in
Eq.~(\ref{123413_28Jul25}).  To compare with conventional spontaneous
symmetry breaking, such as the ferromagnetic transition, it is useful to
introduce a static order parameter. We therefore apply a uniform
external field $h$ to the EOM:
\begin{align}
\dot{\sigma}_i(t) = -\mu(t)\sigma_i(t) + \xi_i(t) + h,
\end{align}
and define the conjugate order parameter
\begin{align}
m(t) = \frac{1}{N}\sum_{i=1}^N \sigma_i(t).
\end{align}
Its equation of motion is
\begin{align}
\dot{m}(t) = -\mu(t)m(t) + h. 
\end{align}
In the steady state, $\dot{m}=0$, yielding 
\begin{align}
m = \frac{h}{\mu}.\label{112958_27Jul25} 
\end{align}
Repeating the analysis of previous sections, the self-consistent
equation for $\mu$ becomes
\begin{align}
 \mu =
 \frac{2T\mu}{\pi}\int_0^\infty d\omega
 \frac{\tD(\omega)}{\mu^2+\omega^2} + hm.\label{215100_30Jul25}
\end{align}
In the limit $h\to 0$, Eqs.~(\ref{112958_27Jul25})
and (\ref{215100_30Jul25}) lead to
\begin{align}
\lim_{h\to \pm 0}m &=
 \pm \sqrt{1-\frac{2T}{\pi}\int_0^\infty d\omega 
\frac{\tD(\omega)}{\omega^2+\mu^2}}\new 
&= 
\begin{cases}
\pm \left(1-\frac{T}{T_c}\right)^\beta & T<T_c \\ 
0 & T\geq T_c
\end{cases},
\end{align}
with the critical exponent $\beta=1/2$, identical to the mean-field
ferromagnetic transition~\cite{nishimori2010elements}. Using
Eqs.~(\ref{112958_27Jul25}) and (\ref{215100_30Jul25}), $m$ can also be
calculated for finite $h$. Fig.~\ref{173331_28Jul25} shows the $h$
dependence $m$ for the high-pass filtered noise. For $T\geq T_c$, $m$
increases continuously with $h$, whereas for $T<T_c$, $m$ changes
discontinuously at $h=0$. Unlike the Ising model, no hysteresis is
observed even for $T<T_c$, because the continuous spin variables can
rotate smoothly toward the external field without overcoming an energy
barrier. This absence of hysteresis has also been reported in the
mean-field Heisenberg model~\cite{shukla2010}.

At the transition point $T=T_c$, $m$ exhibits a singular behavior 
for $\abs{h}\ll 1$:
\begin{align}
m\propto \abs{h}^{\frac{1}{\delta}}.
\end{align}
For the high-pass filtered noise~(\ref{124131_21Jul25}), 
asymptotic analysis of
Eqs.~(\ref{112958_27Jul25}) and (\ref{215100_30Jul25})
yields 
\begin{align}
\delta = 3,
\end{align}
again consistent with the mean-field 
ferromagnetic transition~\cite{nishimori2010elements}.  For the 
power-law spectrum~(\ref{190802_18Jul25}), the critical
exponent $\delta$ depends on $\alpha$. Asymptotic analysis of
Eq.~(\ref{215100_30Jul25}) gives 
\begin{align}
\delta =
 \begin{cases}
  \frac{\alpha+1}{\alpha-1} & \alpha <3 \\
  2& \alpha>3
 \end{cases}.
\end{align}
Thus, $\alpha_{\rm up}=3$ plays the role of the upper critical
dimension, above which the critical exponent saturates to a constant
value.  The value of $\alpha_{\rm up}$ is consistent with
that obtained in the previous section.

\section{Related models}
\label{183622_29Jul25} To examine the generality of interaction-free
ergodicity breaking, we now investigate related models with alternative
types of global constraints, namely, $L_p$ norm constraints and soft
constraints.

\subsection{$L_p$ norm constraint}
We first consider the $L_p$ norm constraint:
\begin{align}
\sum_{i=1}^N \abs{\sigma_i}^p = N.\label{223316_24Jul25}
\end{align}
To enforce this constraint, we renormalize $\sigma_i$ at each time step
according to
\begin{align}
 \sigma_i(t+\Delta t) =
\frac{\sigma_i(t)+\xi_i(t)\Delta t}
{\left(N^{-1}\sum_{i=1}^N\abs{\sigma_i+\xi_i(t)\Delta t}^p\right)^{\frac{1}{p}}}.
\end{align}
In the limit of small $\Delta t$, this yields 
\begin{align}
\dot{\sigma}_i(t) = -\mu(t) \sigma_i(t) + \xi_i(t),\label{223854_24Jul25}
\end{align}
where 
\begin{align}
 \mu(t) = \frac{1}{N}\sum_{i=1}^N \abs{\sigma_i}^{p-2}\sigma_i(t)\xi_i(t).
\end{align}
In the steady state, $\mu$ becomes constant.  Since $\xi_i(t)$ is
Gaussian and Eq.~(\ref{223854_24Jul25}) is linear, the stationary
distribution in the ergodic phase is also Gaussian.  In the ergodic
phase $\ave{\sigma_i}=0$, and then, the distribution is given by
\begin{align}
P(\sigma_i) =
 \frac{1}{\sqrt{2\pi\ave{\sigma_i^2}}}
 e^{-\frac{\sigma_i^2}{2\ave{\sigma_i^2}}},\label{141827_25Jul25}
\end{align}
where 
\begin{align}
\ave{\sigma_i^2} = \frac{2T}{\pi}\int_{0}^\infty d\omega \frac{\tD(\omega)}{\omega^2+\mu^2}.
\end{align}
From this distribution, we obtain
\begin{align}
\ave{\abs{\sigma_i}^p} = 
\int_{-\infty}^{\infty}d\sigma_i P(\sigma_i)\abs{\sigma_i}^p
= A_p\ave{\sigma_i^2}^{\frac{p}{2}},
\end{align}
where
\begin{align}
A_p = \Gamma\left[\frac{1+p}{2}\right]
 \frac{2^{\frac{p}{2}}}{\pi^\frac{1}{2}}.
\end{align}
The constraint~(\ref{223316_24Jul25}) then becomes 
\begin{align}
1 = A_p\left[\frac{2T}{\pi}\int_0^\infty d\omega
\frac{\tD(\omega)}{\omega^2+\mu^2}\right]^{\frac{p}{2}}.\label{230519_24Jul25}
\end{align}
As before, the transition point is obtained by
substituting $\mu=0$ into Eq.~(\ref{230519_24Jul25}), 
leading to 
\begin{align}
T_c &= A_p^{-\frac{2}{p}}\frac{\pi}{2}\left[\int_0^\infty d\omega \frac{\tD(\omega)}{\omega^2}\right]^{-1}.\label{152431_25Jul25}
\end{align}
For the power-law spectrum~(\ref{190802_18Jul25}),
this reduces to
\begin{align}
T_c = 
\begin{cases}
A_p^{-p/2}\frac{\pi\omega_0}{2}\frac{\alpha-1}{\alpha} & \alpha > 1 \\
 0 & \alpha \leq 1 
\end{cases}.
\label{152418_25Jul25}
\end{align}
Thus, when the noise is class-I hyperuniform ($\alpha>1$), the model
undergoes an ergodicity-breaking transition at finite $T_c$, as in the
case of the model with the $L_2$ norm constraint.

\subsection{Soft constraints}
We next consider a model with a soft global constraint defined by 
\begin{align}
\dot{\sigma}_i(t) = \gamma \sigma_i(t)
\left(1-\frac{1}{N}\sum_{j=1}^N\abs{\sigma_j}^p\right) +
\xi_i(t).\label{144510_25Jul25}
\end{align}
For small $\sigma_i$, the growth rate is $\gamma$, while for large
$\sigma_i$, the growth is suppressed by the global feedback term
${1-\sum_j \abs{\sigma_j}^p/N}$, as in the case of the logistic
growth~\cite{murray2007mathematical}. As in Eq.~(\ref{141827_25Jul25}),
the steady-state distribution in the ergodic phase is Gaussian with
variance
\begin{align}
\ave{\sigma_i^2} =
 \frac{2T}{\pi}\int_{0}^\infty d\omega
 \frac{\tD(\omega)}{\omega^2+\gamma^2(1-q)^2},
\end{align}
where 
\begin{align}
q = \frac{1}{N}\sum_{i=1}^N \abs{\sigma_i}^p
 = \int d\sigma P(\sigma)\abs{\sigma}^p.\label{003742_13Aug25}
\end{align}
Unlike the strict-constraint models, 
here $q$ is not fixed to unity
but is determined self-consistently
from Eq.~(\ref{003742_13Aug25}). The restoring force 
in Eq.~(\ref{144510_25Jul25}) vanishes when $q=1$, 
which is the signature of the
ergodicity breaking transition.
The critical point $T_c$ 
is calculated by substituting $q=1$ into Eq.~(\ref{003742_13Aug25}),
leading to 
\begin{align}
 1 = A_p \left(\frac{2T_c}{\pi}\int_0^\infty d\omega
\frac{\tD(\omega)}{\omega^2}
  \right)^{\frac{p}{2}}.
\end{align}
Solving for $T_c$, we find that $T_c$ coincides with that of the
$L_p$ norm conserving model, Eq.~(\ref{152431_25Jul25}). Therefore,
ergodicity breaking occurs not only under strict constraints, but also
for models with smooth global constraints.

\section{Summary and discussions}
\label{164632_2Sep25}

In this work, we have investigated spin models with global constraints
driven by temporally hyperuniform noise. We showed that the system
undergoes a sharp ergodicity-breaking transition at a critical noise
strength $T_c$ whenever the noise belongs to class-I hyperuniformity,
{\it i.e.}, when the power spectrum $\tilde{D}(\omega)$ vanishes at
small frequency as $\tilde{D}(\omega)\propto |\omega|^\alpha$ with
$\alpha>1$.  The central message is that global constraints combined
with strongly anticorrelated driving forces are sufficient to induce
ergodicity breaking, even in the complete absence of explicit
interactions.

An important open question is whether such interaction-free ergodicity
breaking can also emerge in more realistic settings. The key ingredients
are global constraints and class-I hyperuniform driving.  The global
constraints appear in diverse contexts including constraint satisfaction
problems~\cite{mezard2009information,franz2017universality}, theoretical
ecology~\cite{diederich1989,biscari1995,azaele2016,altieri2022glassy},
economics and
finance~\cite{bouchaud2003theory,burda2002,bouchaud2023application}, and
random lasers~\cite{ghofraniha2015,niedda2023}. Moreover, there are
several examples of temporally hyperuniform noise, such as fractional
Brownian motion~\cite{mandelbrot1968fractional}, stochastic
resetting~\cite{evans2011,evans2020}, and high-pass filtered
signals~\cite{guz1998,guz2001,bao2005,hu2017,chen2018,li2021}, see
Appendix~\ref{193904_26Jul25}. A promising direction for future work is
therefore to explore whether, in experimentally or practically relevant
settings, the combination of these factors can indeed drive
interaction-free ergodicity breaking as predicted by our model.

\begin{acknowledgments}
I acknowledge the use of OpenAI’s ChatGPT (https://chat.openai.com/) for
assistance in improving the clarity, grammar, and overall readability of
the manuscript. This project has received JSPS KAKENHI Grant Numbers
23K13031 and 25H01401.
\end{acknowledgments}

\appendix 

\section{Examples of temporally hyperuniform noise}
\label{193904_26Jul25} 
Here we list several examples of temporally hyperuniform noise.

\subsection{High-pass filter}
\label{highpass} Class-I hyperuniform noise can be generated by a
high-pass filter, which suppresses the low frequency components of an
input signal. The simplest realization is the RC circuit consisting of a
resistor and capacitor. From Kirchhoff's first law, the equation of
motion of the circuit is written
as~\cite{guz1998,bao1999,guz2001}
\begin{align}
&RI(t) + \frac{Q(t)}{C} + \eta(t)=0,\new
& I(t) = \dot{Q}(t),\label{225655_10Jul25}
\end{align}
where $Q(t)$, $I(t)$, $R$, and $C$ respectively denote the charge,
current, resistance, and capacitance, and $\eta(t)$ represents the
input white noise satisfying
\begin{align}
\ave{\eta(t)} = 0,\quad
\ave{\eta(t)\eta(t')} = 2T\delta(t-t').\label{135149_1Sep25} 
\end{align}
The output is the voltage across the resistor:
\begin{align}
\xi(t) = RI(t).
\end{align}
A straightforward calculation in the steady state
yields~\cite{guz1998,bao1999,guz2001}
\begin{align}
\int dt e^{i\omega t}\ave{\xi(t)\xi(0)}
 = 2T \tD(\omega)
\end{align}
with 
\begin{align}
\tD(\omega) = \frac{\omega^2}{\omega^2+\left(\frac{1}{RC}\right)^2}.\label{172707_1Sep25}
\end{align}
For small $\omega$, the power spectrum exhibits hyperuniformity:
${\tD(\omega)\propto
\omega^\alpha}$ with $\alpha=2$.

Cascading multiple high-pass filters produces stronger suppression of
the low-frequency components. The power spectrum of the noise after passing
through $n$ such filters is given by 
\begin{align}
 \tD_n(\omega) = 
 \left(\frac{\omega^2}{\omega^2+\left(\frac{1}{RC}\right)^2}\right)^n.
\end{align}
For small $\omega$,
we obtain $\tD_n(\omega)\propto \omega^\alpha$ with
$\alpha=2n$ for small $\omega$.

\subsection{Velocity fluctuations of a Brownian particle confined in a harmonic potential}
\label{harmonic}

The velocity fluctuations of a Brownian particle also exhibit temporal
hyperuniformity~\cite{bao2005,hu2017,chen2018,li2021}. Consider 
an overdamped Brownian particle in a harmonic
potential:
\begin{align}
\dot{x}(t)  = -k x(t) + \eta(t),
\end{align}
where $\eta(t)$ denotes thermal white noise satisfying
Eqs.~(\ref{135149_1Sep25}). In the steady state, the Fourier spectrum
of the velocity $v(t)\equiv \dot{x}(t)$ is given by 
\begin{align}
 \int_{-\infty}^{\infty}dt e^{i\omega t}\ave{v(t)v(0)}
 = 2T\tD(\omega)
\end{align}
with 
\begin{align}
\tD(\omega)&= \frac{\omega^2}{\omega^2+k^2}.\label{172657_1Sep25}
\end{align}
For $\omega\ll 1$, the correlation scales as ${\tD(\omega)\propto
\omega^2}$. Therefore, the velocity fluctuations of a Brownian particle
confined in a harmonic potential are temporally
hyperuniform~\cite{bao2005,hu2017,chen2018,li2021}.  When $k=1/(RC)$,
Eq.~(\ref{172657_1Sep25}) coincides with Eq.~(\ref{172707_1Sep25}),
indicating that the velocity fluctuations follow the same statistics as
the high-pass filtered noise.

\subsection{Fractional Brownian motion}
The position $x(t)$ of fractional Brownian motion (FBM) is a
stochastic variable with zero mean and
variance~\cite{mandelbrot1968fractional}:
\begin{align}
\ave{x(t)x(t')}  = 
 \left[\abs{t}^{2H}+\abs{t'}^{2H}
 -\abs{t-t'}^{2H}\right],
\end{align}
where $H\in (0,1)$
is the Hurst exponent. The FBM can be rewritten in the form of a stochastic
differential equation:
\begin{align}
 \dot{x}(t) = \xi(t),
\end{align}
where the correlation of the noise is 
\begin{align}
D(t-t') &\equiv \ave{\xi(t)\xi(t')} \new 
&= \pdiff{^2\ave{x(t)x(t')}}{t\partial t'}\new 
&=  H(2H-1)\abs{t-t'}^{2H-2}.
\end{align}
The power spectrum of the noise is then
given by~\cite{mandelbrot1968fractional}
\begin{align}
 \tD(\omega) = \int dt e^{i\omega t}D(t)
\sim \abs{\omega}^\alpha,\ \alpha = 1-2H.
\end{align}
For $\alpha>0$ or equivalently $H<1/2$, $\tD(\omega)$ vanishes for small
$\omega$, implying that the noise is temporally hyperuniform.

\subsection{Brownian motion with stochastic resetting}
We next consider Brownian motion with stochastic resetting, where the
position $x(t)$ of a Brownian particle is reset to its initial value 
$x_0$ at a constant rate $r$~\cite{evans2011}. The update rule
for a small time increment $\Delta t$ 
can be written as
\begin{align}
 x(t+\Delta t) = \begin{cases}
	    x_0 & {\rm probability}\ r \Delta t \\
	    x(t)+ W_t\sqrt{2T\Delta t}& {\rm probability}\ 1-r \Delta t
	   \end{cases},
\end{align}
where $W_t$ denotes a random number
satisfying
\begin{align}
 \ave{W_t} =0,\quad \ave{W_tW_t'} = \delta_{t,t'}.
\end{align}
To simplify the notation, we hereafter rescale the units such that the
noise strength is normalized to unity, {\it i.e.}, $2T \to 1$. Since
the resets follow a Poisson process, the waiting time $\tau$ between
successive resets is exponentially distributed~\cite{evans2020}:
\begin{align}
 \psi(\tau) = re^{-r\tau}.\label{181127_10Jul25}
\end{align}
In the interval between two consecutive resets $t_i<t\leq t_i+\tau$,
the equation of motion in the continuum limit $\Delta t\to 0$ 
can be formally written as
\begin{align}
\dot{x}(t) = \xi(t),
\end{align}
with 
\begin{align}
\xi(t) &\equiv \lim_{\Delta t\to0 }
 \frac{x(t+\Delta t)-x(t)}{\Delta t}\new
 &= \eta(t) -\delta(t-t_i-\tau)\int_{t_i}^{t_i+\tau}dt\eta(t),
\end{align}
where $\eta(t)$ denotes a white noise with variance
$\ave{\eta(t)\eta(t')}=\delta(t-t')$, and the second term accounts for
the effect of the reset at $t=t_i+\tau$. The Fourier transform of the
noise in this interval is
\begin{align}
\tilde{\xi}_i(\omega) &\equiv \int_{t_i}^{t_i+\tau}dt e^{i\omega t}\xi(t)\new 
& = \int_{t_i}^{t_i+\tau}\left(e^{i\omega t}-e^{i\omega(t_i+\tau)}\right)
 \eta(t).
\end{align}
The corresponding power spectrum depends only on $\tau$ and not on $i$:
\begin{align}
\tD_{\tau}(\omega) &= \ave{\abs{\xi_i(\omega)}^2}\new
 &= 2\tau \left[1-\frac{\sin(\omega\tau)}{\omega \tau} \right].
\end{align}
Because $\xi_i(\omega)$ and $\xi_j(\omega)$ are statistically
independent for $i\neq j$,
the total spectrum can be obtained 
by summing the contributions from each interval:
\begin{align}
\tD(\omega) &= \lim_{\T\to\infty}\frac{1}{\T}
 \abs{\int_0^{\T} e^{i\omega t}\xi(t)}^2\new 
& =
 \lim_{\T\to\infty}\frac{1}{\T}
 \sum_{i=1}^{N_{\T}}\ave{\abs{\tilde{\xi}_i(\omega)}^2},
\end{align}
where $N_{\T}$ denotes the number of resets during the interval
$\T$. After some manipulations, this expression reduces to 
\begin{align}
\tD(\omega) = \frac{\ave{\tD_\tau(\omega)}}{\ave{\tau}},\label{183400_10Jul25} 
\end{align}
with 
\begin{align}
\ave{\tD_{\tau}(\omega)} &= \lim_{\T\to\infty}\frac{1}{N_\T}\sum_{i=1}^{N_\T}
\ave{\tilde{\xi}_i(\omega)^2}\new 
&= \int_0^\infty d\tau \psi(\tau)\tD_\tau(\omega),
\end{align}
and
\begin{align}
 \ave{\tau} =
\lim_{\T\to\infty}\frac{\T}{N_\T}
= \int_0^\infty d\tau \psi(\tau)\tau.
\end{align}
For the exponential distribution Eq.~(\ref{181127_10Jul25}), we finally
obtain 
\begin{align}
 \tD(\omega) = \frac{2\omega^2}{r^2+\omega^2}.
\end{align}
For small $\omega$, the spectrum exhibits a power-law scaling
$\tD(\omega)\propto \omega^\alpha$ with $\alpha=2$, which is the
signature of temporal hyperuniformity. The spectrum of the positional
correlation $\tilde{C}(\omega)\equiv \int dt e^{i\omega
t}\ave{x(t)x(0)}$ is calculated as
$\tilde{C}(\omega)=\tD(\omega)/\omega^2=2/(r^2+\omega^2)$, which is
consistent with previous work~\cite{majumdar2018}.

The result can be generalized to non-Poisson resetting, which leads to
different values of $\alpha$. As an example, consider the
power-law waiting time distribution~\cite{bodrova2019}:
\begin{align}
 \psi(\tau) = \frac{\beta}{\tau_0}\left(\frac{\tau}{\tau_0}\right)^{-\beta-1}
\quad \tau>\tau_0,\label{183343_10Jul25}
\end{align}
where $\beta>1$ is required to keep the mean waiting time $\ave{\tau}$
finite.  Substituting Eq.~(\ref{183343_10Jul25}) into
Eq.~(\ref{183400_10Jul25}) and performing an asymptotic analysis for
$\omega\ll 1$, we obtain 
\begin{align}
\tD(\omega) \propto \abs{\omega}^\alpha,
\end{align}
with 
\begin{align}
\alpha =
 \begin{cases}
  \beta-1 & 1<\beta<3 \\ 
  2& \beta >3
 \end{cases}.	  
\end{align}
The spectrum thus exhibits class-I hyperuniformity for $\beta>2$,
class-II hyperuniformity for $\beta=2$, and class-III hyperuniformity for
$\beta<2$.

\subsection{One-dimensional hyperuniform sequences}

There are several examples of hyperuniform one-dimensional sequences,
including quasicrystals and substitution tilings~\cite{oguz2017}, as
well as one-dimensional lattices with stochastic particle
displacements~\cite{kim2018effect}. These can also be regarded as
examples of temporally hyperuniform sequences when the spatial
coordinate $x$ is interpreted as time $t$.

\section{A free Brownian particle driven by temporally hyperuniform noise}
\label{180030_11Sep25}
We analyze the Brownian motion driven by
the power-law noise in one
dimension. The dynamics is governed by the Langevin equation:
\begin{align}
 \dot{x}(t) = \xi(t),\label{161819_11Jul25}
\end{align}
where $T$ denotes the noise strength, 
and $\xi(t)$ is a stochastic process 
with zero mean 
and variance:
\begin{align}
\ave{\xi(t)\xi(t')} = 2TD(t).
\end{align}
We assume the following Fourier spectrum of $D(t)$:
\begin{align}
\tD(\omega) \equiv \int_{-\infty}^{\infty} dt e^{i\omega t}D(t) 
 =
 \begin{cases}
  \abs{\omega/\omega_0}^\alpha & \abs{\omega}\leq \omega_0 \\ 
  1 & \abs{\omega}> \omega_0
 \end{cases}.
\end{align}
Thermal white noise is recovered for $\alpha=0$, while FBM corresponds
to the limit $\omega_0\to\infty$~\cite{mandelbrot1968fractional}. The
finite cut-off $\omega_0$ enables convergence of the MSD even for
$\alpha>1$, as shown below. The MSD is obtained by integrating
Eq.~(\ref{161819_11Jul25}):
\begin{align}
{\rm MSD} &\equiv \ave{\left(x(t)-x(0)\right)^2} =
 \int_0^t dt_1 dt_2 \ave{\xi(t_1)\xi(t_2)}\new
&= \frac{4T}{\pi}\int_0^\infty d\omega  
\frac{1-\cos(\omega t)}{\omega^2}\tD(\omega).
\end{align}
An asymptotic
analysis for large $t$ yields
\begin{align}
{\rm MSD} \propto 
 \begin{cases}
  t^{1-\alpha} & \alpha < 1\\ 
  \log t  & \alpha = 1 \\ 
  const & \alpha >1.
 \end{cases}
\end{align}
The MSD grows sublinearly for $0<\alpha<1$, logarithmically for
$\alpha=1$, and saturates at long times for $\alpha>1$.

\bibliography{reference}
\end{document}